\documentstyle[12pt]{article} 
 1                                        
\begin{document}                                                              
%\begin{flushright}{LNF-94/xxx P}  
%\end{flushright}                                                      

\begin{center}                                                                
{ \large\bf 
LONG RANGE FORCES FROM QUANTUM FIELD THEORY AT ZERO AND FINITE TEMPERATURE}

\vskip 2cm  
{M. Nowakowski}\\
Departamento de Fisica, Universidad de los Andes, A.A. 4976, Santafe de Bogota, D.C.,
Colombia \\
E-mail: {marek@marik.uniandes.edu.co}

\end{center}

\vskip .5cm                             
\begin{abstract} 
We discuss the derivation of Newtonian potentials in the framework of quantum field
theory. We focus on two particular points: on 
long range forces i.e. forces which fall off as $1/r^n$
being mediated by light quanta like neutrinos or Goldstone bosons and on possible
temperature dependence of such forces arising in situations when the exchanged quanta are in 
a thermal heat bath. Examples of the latter are cosmic relic photons and relic neutrinos.
Among other things, we will show that the existence of cosmic relic neutrinos modifies the
long tail of the two-neutrino exchange Feinberg-Sucher force drastically. Results concerning
the potential mediated by two Goldstone bosons are also presented.  

\end{abstract}                                                                

\newpage
\section{Overview}
One of the most important key concepts in theoretical physics, is the concept of a force
introduced by Newton some three hundred years ago. Without any doubts, this concept continues to play
a fruitful role in physics, despite the fact that classical mechanics has been superseded by the 
more general quantum theory. Indeed, modern theories of interactions use the tools of quantum
field theory (QFT) as a general framework. It bears therefore a certain 
charm when we can span a
bridge between classical mechanics and QFT by deriving new forces, 
especially the long range forces,
within the framework of the latter. However, this is not the only reason which makes the subject
worthwhile as the following rough classification of long range forces demonstrates.\\
1. \underline{Quantum corrections to classical results.}
The QFT can, of course, reproduce the classical long range forces of electromagnetism and gravity.
In addition QFT predicts also quantum mechanical corrections to these classical results. So,
for instance the Coulomb potential receives corrections of the following type
\cite{landau}
\begin{eqnarray} \label{coulomb}
V_{em}(r)&=&{e^2 \over r}\left[1 + \delta V^{QM}_{em}(r)\right]
\nonumber \\
\delta V^{QM}_{em}(r)&=& {2\alpha \over 3\pi}\left (\ln (1/m_e r)-C -{5 \over 6}\right )
\nonumber \\
&-&{2\alpha^2 e^2 \over 225 \pi}{1 \over (m_e r)^4}
\end{eqnarray}
where $C=0.577$ and $m_e$ is the electron mass. The first correction ($\sim \alpha$)
is due to vacuum polarization and valid for $m_er \ll 1$. The second correction ($\sim \alpha^2$)
has its root in the Heisenberg-Euler Lagrangian ($\gamma \gamma$ scattering). Similarly, for gravity
using low energy effective field theory techniques, one derives quantum corrections of the form
\cite{donoghue} 
\begin{eqnarray} \label{gravity}
V_{gravity}(r)&=&-{G_NM_1M_2 \over r}\left[1 + \delta V^{QM}_{gravity}(r)\right]
\nonumber \\
\delta V^{QM}_{gravity}(r)&=& -{G_N(M_1+M_2) \over r}
+{127G_N \over 30\pi^2r^2} \nonumber \\
\end{eqnarray}
2. \underline{New forces from old quanta.} 
All forces in the QFT arise from the exchange of quanta,
massless or very light in the case of long range forces.
Apart from the corrections to the classical results given above, 
the next logical step is to search
for possible long range forces mediated by some light particles in the experimentally established
particle spectrum. Neutrinos, being either very light or massless, are the natural candidates.
This was suspected by Feynman \cite{feynman} 
and demonstrated in detail by Feinberg and Sucher \cite{FS}. 
We quote below
the more general results for massive neutrinos distinguishing between Dirac and Majorana type
\cite{mFS}
\begin{eqnarray} \label{Tzero}
V_{Dirac}(r)&=&{G_F^2m_{\nu}^3g_{V}g_{V}'\over 4\pi^3r^2}K_3(2m_{\nu}r)
\nonumber \\
V_{Majorana}(r)&=&{G_{F}^2m_{\nu}^2g_{V}g_{V}'\over 2\pi^3r^3}K_{2}(2m_{\nu}r)
\nonumber \\
\end{eqnarray}
where $g_V$ and $g_{V}'$ are vector coupling constants and $K_n$ are modified Bessel functions.
For $m_{\nu}=0$ both results reduce to the formula obtained originally by Feinberg and Sucher,
namely
\begin{equation}\label{FSR}
V_{FS}(r)={G_{F}^2g_{V}g_{V}'\over 4\pi^3r^5}.
\end{equation}
Another famous example is the Casimir-Polder force mediated by two photons between polarizable
particles \cite{CP}. Its analytical form reads
\begin{equation}\label{casimirpolder}
V_{CP}(r)=-{23(\alpha_E^2+\alpha_B^2) -14(\alpha_E\alpha_B) \over (4\pi)^3 r^7}
\end{equation}
where $\alpha_E$ and $\alpha_B$ are electric and magnetic polarizabilities of the external particle.
It is worthwhile mentioning that it took some fifty years to verify this force experimentally
\cite{expCP}. This also shows that a technologically difficult task, not possible at the moment, 
could still become feasible in the future. 
We emphasize this, because all forces we are discussing here are feeble and difficult to detect
experimentally. \\
3.\underline{New forces from new quanta.} 
While going beyond the Standard Model we can, in principle,
encounter many other light quanta, mostly light  
scalars, pseudo-scalars or true Goldstone bosons.
Famous examples are Axions \cite{axion}, Majorons 
\cite{majoron} and scalars and pseudo-scalars in no-scale supergravity 
\cite{noscale} and others \cite{fayet}.
A tower of massive gravitons is also possible by compactification of extra higher 
space dimensions \cite{gravitons}. 
Hence a search for long range forces mediated by such exotic particles
could be a harbinger of new physics. For more details, especially regarding the experimental aspect 
of such a search, we refer the reader to \cite{exp}. \\
4. \underline{New effects: temperature dependent forces.}\\ 
From the point of view of QFT at finite 
temperature, an exchange of quanta which are in a thermal bath at a temperature $T$
leads, of course, to temperature dependent amplitudes and therefore also to temperature dependent
forces. This is indeed a curious prediction of QFT at finite temperature. Physical examples of quanta in 
a thermal heat bath are cosmic relic photons (microwave background radiation) and relic neutrinos
(the latter not yet experimentally verified). 
In the real time approach to finite temperature field theory the full propagator is a matrix 
out of which we need for the actual calculations of potentials only the 1-1 component
given by
\begin{eqnarray}\label{propagator}
&&S_T^{fermion}(k)=(\rlap / k +m)[(k^2-m^2+i\epsilon)^{-1}
\nonumber \\
&&+2\pi i\delta(k^2-m^2)(\theta
(k^0)n_{+}(T)+\theta(-k^0)n_{-}(T))]
\nonumber \\
&&S_T^{boson}(k)=(k^2-m^2+i\epsilon)^{-1}
\nonumber \\
&&-2\pi i\delta(k^2-m^2)(\theta
(k^0)n_{+}(T)+\theta(-k^0)n_{-}(T))\nonumber \\
\end{eqnarray}
where $n_{+}$ and $n_{-}$ are distribution functions for
particle and antiparticle, respectively. 
Temperature corrections to various long range forces have been calculated in
\cite{HP, we1, we2, we3, FG1, FG2}.
\\
5. \underline{Forces not derivable from QFT.} 
We mention here for completeness that an example of such a force would be the Newtonian
limit of Einstein's gravity with a cosmological constant $\Lambda$. For a spherical object
or point-like particle, the gravitational potential reads
\begin{equation}\label{lambda}
\Phi_{\Lambda}(r)=-{G_N M \over r} -{1 \over 6}\Lambda r^2
\end{equation}
The second part, proportional to $\Lambda$, cannot be derived from QFT. 
If certain recent experimental indications of a non-zero cosmological constant should be confirmed,
the $\Lambda$-force in the Newtonian approximation would be ``longest'' out of the long range 
forces in nature.
For peculiarities of the Newtonian
limit in the presence of non-zero $\Lambda$, see \cite{melambda}.

Before discussing concrete examples, a few comments about the actual method to calculate a potential
from an amplitude ${\cal M}$ are in order. There are essentially 2 equivalent methods.  
The more
standard one is to take the Fourier transform of a matrix element in the static limit i.e.
approximating the four momentum transfer $q$ by $q \simeq (0, {\mathbf Q})$ ($Q=\vert {\mathbf Q}
\vert$).
\begin{eqnarray} \label{potdef2}
V(r) &=& \int {d^3 Q \over (2\pi)^2} \exp(i{\mathbf Q}{\mathbf r}){\cal M}
({\mathbf Q})
\nonumber \\
&=&{1 \over 2 \pi^2 r}\int^{\infty}_0 dQ Q 
{\cal M}(Q)\sin Qr
\end{eqnarray}
The other, more elaborate, method uses dispersion techniques and defines \cite{FSA}
\begin{equation} \label{defpot}
V(r)={-i \over 8\pi^2 r}\int^{\infty}_{4m^2}
dt [{\cal M}]_t\exp (-\sqrt{t}r)
\end{equation}
where the integration variable $t$ 
 equals the four--momentum transfer
squared, $q^2$. Here, 
$[{\cal M}]_t$ denotes the discontinuity
of the Feynman amplitude across the cut
in the real $t$ axis.

In the next sections we will focus on two particular examples with slightly different emphasis.
The first example will be the two-neutrino exchange force (Feinberg-Sucher force). 
We will examine here
the aforementioned temperature dependence taking different thermal distributions 
$n_{\pm}(T)$. The second
example deals with the two-boson exchange force and emphasizes the difference between light
pseudoscalar and Goldstone bosons.

\section{Two-neutrino exchange force} 
Given (\ref{potdef2}), the potential follows once we
have calculated the matrix element ${\cal M}$ using (\ref{propagator}). Let us start with a
simple example of classical Boltzmann-distribution.
\\
a.\underline{Boltzmann-distribution:
$n_{\pm}=e^{[(\pm \mu -\vert k^0\vert)/T]}$.}\\
With this distribution, the integrations involved in the calculation of potentials can
be easily done by conveniently choosing the order in which they are performed. The
results can be expressed again in terms of Bessel functions and read \cite{we1}:
\begin{eqnarray}\label{diracT}
V_{T}^{Dirac}(r)&=&-{G_{F}^2m_{\nu}^4g_{V}g_{V}'\over \pi^3r}\cosh {(\mu/T)}
\nonumber \\
&\times &\left [{K_{1}(\rho)\over \rho}+{4K_{2}(\rho)\over \rho^2} \right]
\end{eqnarray}
and
\begin{equation}\label{majoT}
V_{T}^{Majorana}(r)=-{4G_{F}^2m_{\nu}^4g_{V}g_{V}'\over \pi^3r}{K_{2}(\rho)\over \rho^2}
\end {equation}
where we have defined 
\begin{equation}
\rho\equiv {m_{\nu}\over T}\sqrt{1+(2rT)^2}.
\end{equation}
For massless neutrinos (and $\mu=0$) both potentials collapse to
\begin{equation}\label{zeromT}
V_{T}(r)=-{8G_{F}^2m^4g_{V}g_{V}'\over \pi^3r}{1\over \rho^4}
\end {equation}
which is the result given in reference \cite{HP}.
We see that for
distances much larger than $T^{-1}$ the potential reads
\begin{equation}\label{asympt}
V_{T}(r)\simeq -{G_{F}^2g_{V}g_{V}'\over 2\pi^3r^5}.
\end{equation}
When added to the vacuum result (\ref{FSR}), the total potential is
\begin{equation}\label{totalV}
V_{tot}(r)\simeq -{G_{F}^2g_{V}g_{V}'\over 4\pi^3r^5}
\end {equation} 
that is, in the presence of a thermal neutrino background,
distributed according to the Boltzmann distribution, the original Feinberg-Sucher
force switches sign, i.e. a repulsive force turns into an attractive one. On the other
hand, for ($rT\ll 1$), the temperature dependent
potential (16) behaves as follows
\begin{equation}\label{submm}
V_{T}(r)\simeq -{8G_{F}^2g_{V}g_{V}'T^4\over \pi^3r}
\end{equation}
which is negligible compared to the vacuum contribution in equation (\ref{FSR}). 
\\
b.\underline{Cold degenerate neutrinos: $n_+=\theta (\mu -k^0)$}.\\
The main interest in such distributions is the physics of supernova.
Here we find for the potential (assuming $m_{\nu}=0$)
\begin{equation}\label{Tpot11}
V_{T}(r)\simeq -2V_{FS}(r)[1-\cos 2\mu r-\mu r\,\sin
2\mu r]
\end{equation}
which agrees with the result given in \cite{HP} and in \cite{SV, we2}.
\\
c.\underline{Fermi-Dirac: 
$n_{\pm}=(e^{(k^0\mp\mu)/T}+1)^{-1}$}\\
The result for $m_{\nu}=0$ can be written in the form \cite{we2}
\begin{equation}\label{Tpot3}
V_{T}(r)=-{G_{F}^2g_{V}g_{V}'\over
4\pi^3r^4}\left[1-r{d\over
dr}\right]I_{T}(r;\mu)
\end {equation}
with the final result being expressible in terms of the
hypergeometric function
$F(a,b;c;z)$. Indeed, we have
\begin{eqnarray}\label{Tpot5}
I_{T}(r;\mu)&=&{1\over
4r}[F(1,-2irT;1-2irT;-e^{-\mu/T})
\nonumber \\
&+&F(1,-2irT;1-2irT;-e^{\mu/T})
\nonumber \\
&+&F(1,2irT;1+2irT;-e^{-\mu/T})
\nonumber \\
&+&F(1,2irT;1+2irT;-e^{\mu/T}) \nonumber \\
&-& 8\pi rT\,\cos 2r\mu\,{\mathrm {csch}}\,2\pi rT], 
\end{eqnarray} 
Let us take nondegenerate
neutrinos ($\mu=0$). 
After some algebra we obtain
$I_{T}(r;\mu=0)$ in the form
\begin{equation}\label{Tpot6}
I_{T}(r;\mu=0)=
{1\over 2r}[1
-2\pi rT\,{\rm csch}\,2\pi rT]
\end{equation}  
such that the temperature dependent potential for
nondegenerate relic
neutrinos is:           
\begin{eqnarray}\label{Tpot7}
&&V_{T}(r)=-V_{FS}(r)\times
\nonumber \\
&&[1-\pi rT\,{\rm
csch}\,2\pi rT(1+2\pi rT\,\coth 2\pi rT)]
\nonumber \\
\end {eqnarray}
where
$V_{FS}(r)$ is the Feinberg-Sucher potential.
At
large distances (i.e. $rT\gg 1$) the temperature dependent effect 
exactly cancels the vacuum
component,  
\begin{equation}\label{Tpot9}
V_{T}(r)\approx -V_{FS}(r)
\end {equation}
This is, indeed, a drastic effect of relic cosmic neutrinos. It makes the long tail
of the Feinberg-Sucher force effectively non-operative. Note that the new scale set by the 
temperature is $T^{-1} \simeq 1mm$. In a supra-millimeter range a future experiment 
searching for the 
Feinberg-Sucher force should give a zero result due to cosmic relic neutrinos!
\section{Two-boson exchange forces}
In the following we will need two generic interactions: one of heavy Higgses (called in the following
$H$) with fermions and of two light or massless pseudoscalars $a$ with the heavy scalars $H$.
We assume that the pseudoscalars do not have tree level coupling to the fermions. Contracting
the heavy Higgs propagator, the Feynman diagram looks formally the same like the diagram
responsible for the Feinberg-Sucher force, of course with the internal 
fermions exchanged by bosons \cite{we3}. For two-boson forces arising from yet different Feynman
diagrams see \cite{grifols}.\\
a.\underline{Light pseudoscalar.}
Consider the case of some generic non-derivative interaction terms of the form
\begin{equation} \label{int.generic}
{\cal L}_{int}=g_{{}_{Hff}}\bar{f}fH, \,\,\,\,\, 
{\cal L}'_{int}=g_{{}_{Haa}}aaH
\end{equation}
where $f$ are standard fermions, $H$ is the heavy Higgs with mass
$m_H$ and $a$ is the very light pseudoscalar with mass $m_a$. 
It is convenient to define global coupling constants as
\begin{equation} \label{GG'}
G(q^2) \equiv {g_{{}_{Hff}}g_{{}_{Haa}} \over q^2 -m_H^2}, \,\,\,\,\,
G'(q^2) \equiv {g_{{}_{Hf'f'}}g_{{}_{Haa}} \over q^2-m_H^2}
\end{equation}
To compute the potential we now use equation (\ref{defpot}) and obtain for the discontinuity 
\begin{eqnarray} \label{calc.Gamma}
[\Gamma]_t&=&\int {{d^4 k \over (2 \pi)^6}
\delta(k^2-m_a^2)\delta(\bar{k}^2-m_a^2)\theta(k^0)\theta(\bar{k}^0)}\nonumber
\\
&=& {1 \over 8\pi}\sqrt{1-\frac{4 m_a^2}{t}}.
\end{eqnarray}
with $[{\cal M}]_t=-i2G(0)G'(0)[\Gamma]_t$ which has to be inserted
into (\ref{defpot}) to compute the final expression \cite{we3}.
\begin{eqnarray} \label{finalpot}
V(r)&=& -\frac{G(0)G'(0)}{4 \pi^2 r}\int^{\infty}_{4m_a^2}{dt 
[\Gamma]_t\exp (-\sqrt{t}r)}
\nonumber \\
&=& -\frac{G(0)G'(0) m_a}{8 \pi^3 r^2} K_1(2m_ar)
\nonumber \\
&\simeq & -{G(0)G'(0) \over 16\pi^3r^3}
\end{eqnarray}
where the last expression is valid for $rm_a \ll 1$.\\
b.\underline{The case of Goldstone bosons.}
It is now convenient to use the following derivative interaction
\begin{equation} \label{interaction2}
{\cal L}''_{int}=\tilde{g}_{{}_{Haa}}H(\partial^{\mu}a)(\partial_{\mu}a).
\end{equation}
We define also
over-all coupling constants $\tilde{G}(q^2)$ and $\tilde{G}'(q^2)$ in analogy to (\ref{GG'}).
As in the preceding case we start with the dispersion 
theoretical definition
of the potential i.e. eq. (\ref{defpot}) where we denote now the matrix
element by $\tilde{{\cal M}}$ given by
\begin{eqnarray} \label{matrixelement2}
\tilde{{\cal M}}&\simeq & -2 i \tilde{G}(0)\tilde{G}'(0)\cdot \tilde{\Gamma} 
\nonumber \\
\tilde{\Gamma}&=&\int{\frac{d^4k}{(2\pi)^4}\frac{i}{k^2}\frac{i}{\bar{k}^2}
(k \cdot \bar{k})^2}
\end{eqnarray}
where as before $\bar{k}=q-k$. 
For the discontinuity we obtain
\begin{eqnarray} \label{discont2}
\left[ \tilde{\Gamma} \right]_t&=&\frac{q^{\mu}q^{\nu}}{(2\pi)^2} \int { d^4k
  \;  \delta(k^2) \delta(\bar{k}^2)\: k_{\mu}k_{\nu}} \nonumber \\
&=&\frac{q^{\mu}q^{\nu}}{(2\pi)^2}\frac{\pi}{2}\left[
\frac{1}{3} \left(q_{\mu}q_{\nu}-\frac{1}{4}g_{\mu \nu} q^2 \right)
  \right]\nonumber\\ 
&=&\frac{t^2}{32 \pi}
\end{eqnarray} 
with $q^2=t$ as usual.
It remains to calculate the integral transform of this discontinuity.
To distinguish the potential from the results in the
preceding section we will call the
potential due to two pseudoscalar exchange arising from the interaction
(\ref{interaction2}), $\tilde{V}$. For the latter we get \cite{we3}
\begin{eqnarray} \label{potderivative}
\tilde{V}(r)&=&-\frac{\tilde{G}(0)\tilde{G}'(0)}{128 \pi^3 r} \int_0^{\infty}
{ dt  \exp (-\sqrt{t}r) t^2} \nonumber \\
&=& -\frac{15\tilde{G}(0)\tilde{G}'(0)}{8 \pi^3 r^7}.
\end{eqnarray}
Had we used a non-derivative coupling scheme for the Goldstone boson interaction with heavy Higgses
we would get in the zeroth order
$(GG')(q^2=0)=0$ and only in the next order $(GG')(q^2)\propto q^4$. This actually means that the 
calculations with the two different coupling schemes yield the same result
which is also a consequence of a general theorem. 
The latter ensures independence of physical results on the
parameterization of the fields \cite{equiv}.


\begin{thebibliography}{99}
\bibitem{landau}
V. B. Berestetskii, E. M. Lifschitz and L. P. Pitaevskii, 
{\it Quantum Electrodynamics},
Butterworth-Heinemann 1982
\bibitem{donoghue} 
J. F. Donoghue, Phys. Rev. Lett. {\bf 72} (1994) 2996
\bibitem{feynman}
R. P. Feynman, F. B. Morinigo and W. G. Wagner, {\it `Feynman Lectures 
on Gravitation'}, Addison-Wesley, Readings, MA 1995
\bibitem{FS}
G. Feinberg and J.
Sucher, Phys. Rev. {\bf A166} (1968) 1638; S. D. H. Hsu and P.
Sikivie,
Phys. Rev.{\bf D49} (1994) 4951
\bibitem{mFS}  
J. A. Grifols, E. Masso and R., Toldra, Phys. Lett. {\bf B389} (1996) 363;
E. Fischbach, Ann. Phys. (N.Y.) {\bf 247} (1996) 213
\bibitem{CP}
H. B. G. Casimir and P. Polder, Phys. Rev. {\bf 73} (1943) 360; 
E. M. Lifschitz, JETP. Lett. {\bf 2} (1956) 73
\bibitem{expCP}
G. I. Sukenik, M. G. Boshier, D. Cho, V. Sandoghdar and E. A. Hinds, 
Phys. Rev. Lett. {\bf 70} (1993) 560.
\bibitem{axion}
R. D. Peccei, T. T. Wu and T. Yanagida,  Phys. Lett. {\bf B172}{1986}{435} and references therein.
\bibitem{majoron}
Y. Chikashige, R. N. Mohapatra and R. D. Peccei, Phys. Lett. {\bf B98} (1981) 265
\bibitem{noscale}
T. Bhatttacharya and P. Roy, Phys. Rev. {\bf D38} (1988) 2284
\bibitem{fayet}
P. Fayet, Class. Quant. Grav.{\bf 13} (1996) A19
\bibitem{gravitons}
S. Dimopoulos, M. Dine, S. Raby and S. Thomas, Phys. Rev. Lett. {\bf 76} (1998) 70; 
I. Antoniadis, S. Dimopoulos and G. Dvali, Nucl. Phys. {\bf B516} (1998) 70
\bibitem{exp}
E. G. Adelberger, B. R. Heckel, C. W. Stubbs and W. F. Rogers,
Ann. Rev. Nucl. Part. Sci. {\bf 41} (1991) 269; 
J. C. Price  in {\it International Symposium on
Experimental Gravitational Physics}, ed. P. F. Michelson, H. Enke and
G. Pizzella, 
World Scientific, 1987; G. Feinberg and J. Sucher in {\it Long Range Casimir Forces: theory and Recent
Experiments in Atomic Systems}, edited by S. F. Levin and D. A. Micha, Plenum, New York 1993;
E. Fischback and C. L. Talmadge, {\it The Search for Non-Newtonian Gravity}, Springer-Verlag, 1999
\bibitem{HP}
C. J. Horowitz and J. Pantaleone, Phys. Lett. {\bf B319} (1993) 186
\bibitem{we1}
F. Ferrer, J. A. Grifols and  M. Nowakowski, Phys. lett. {\bf B446} (1999) 111
\bibitem{we2}
F. Ferrer, J. A. Grifols and M. Nowakowski, Phys. Rev. {\bf D61} (2000) 057304
\bibitem{we3}
F. Ferrer and M. Nowakowski, Phys. Rev. {\bf D59} (1999) 075009
\bibitem{FG1}
F. Ferrer and J. A. Grifols, Phys. Lett. {\bf B460} (1999) 371
\bibitem{FG2}
F. Ferrer and J. A. Grifols, ``{\it Effects of Bose-Einstein Condensation
on Forces among Bodies Sitting in a Heat bath}'', hep-ph/0001185
\bibitem{melambda}
M. Nowakowski, ``{\it The Consistent Newtonian Limit of Einstein's Gravity with a Cosmological
Constant}'', gr-qc/0004037 
\bibitem{FSA}
G. Feinberg, J. Sucher and C.-K. Au, Phys. Rep.{\bf 180} (1989) 83
\bibitem{SV}
A. Yu. Smirnov and F. Vissani, ``{\it Long Range Neutrino Forces and the Lower bound
on Neutrino Mass}'' hep-ph/9604443 
\bibitem{grifols}
J. A. Grifols and S. Tortosa, Phys. Lett. {\bf B328} (1994) 98;
F. Ferrer and J. A. Grifols, Phys. Rev. {\bf D58} (1998) 096006;
\bibitem{equiv}
R. Haag, Phys. Rev. {\bf 112} (1958) 669; S. Coleman, J. Wess and
B. Zumino, Phys. Rev. {\bf 172} (1969) 2239; C. G. Callan, S. Coleman,
J. Wess and B. Zumino, Phys. Rev. {\bf 177} (1969) 2247
\end{thebibliography}
\end{document}